\documentclass{elsart3p}

\usepackage{natbib}

 \usepackage{graphics}
 \usepackage{graphicx}
 \usepackage{subfigure}
 \usepackage{epsfig}
\usepackage{amssymb,amsmath,amsfonts}

 \usepackage{lineno}

\begin{document}

\begin{frontmatter}



 \title{Technology challenges for space interferometry: the option of mid-infrared integrated optics}


 \author[label1,label2]{L. Labadie \corauthref{cor}}
 \ead{labadie@mpia.de}
 \author[label2]{, P. Kern}
 \author[label3]{, P. Labeye}
 \author[label2]{, E. LeCoarer}
 \author[label4]{, C. Vigreux-Bercovici}
 \author[label4]{, A. Pradel}
 \author[label5]{J.-E. Broquin}
 \author[label6]{, V. Kirschner}

 \address[label1]{Max-Planck Institut f\"ur Astronomie, K\"onigstuhl, 17, D-69117, Heidelberg, Germany}
 \address[label2]{Laboratoire d'Astrophysique de Grenoble, BP53, F-38041 Grenoble Cedex 9, France}
 \address[label3]{Laboratoire d'Electronique et des Technologies de l'Information (CEA), 17, rue des Martyrs, 38054 Grenoble Cedex 9, France}
 \address[label4]{Laboratoire de Physico-Chimie de la Mati\`ere Condens\'ee, Institut Gehrardt, Place Eug\`ene Bataillon, 34095 Montpellier Cedex 5, France}
 \address[label5]{Institut de Microelectronique et Electromagn\'etisme et Photonique, 23, rue des Martyrs, 38016 Grenoble Cedex 1, France}
 \address[label6]{ESTEC-ESA, PO Box 299,2200 AG Noordwijk, The Netherlands}

 \corauth[cor]{Corresponding author. Present address: MPIA, K\"onigstuhl, 17,
 D-69117, Heidelberg, Germany. Tel.:+49-6221-528-239}



\begin{abstract}
Nulling interferometry is a technique providing high angular resolution which is the core of the space missions {\it Darwin} and the {\it Terrestrail Planet Finder}. The first objective is to reach a deep degree of starlight cancelation in the range 6 -- 20 $\mu$m, in order to observe and to characterize the signal from an Earth-like planet. Among the numerous technological challenges involved in these missions, the question of the beam combination and wavefront filtering has an important place. A single-mode integrated optics (IO) beam combiner could support both the functions of filtering and the interferometric combination, simplifying the instrumental design. Such a perspective has been explored in this work within the project {\it Integrated Optics for Darwin} (IODA), which aims at developing a first IO combiner in the mid-infrared. The solutions reviewed here to manufacture the combiner are based on infrared dielectric materials on one side, and on metallic conductive waveguides on the other side. With this work, additional inputs are offered to pursue the investigation on mid-infrared photonics devices.
\end{abstract}

\begin{keyword}
Space-based nulling interferometry \sep Single-mode integrated optics \sep mid-infrared instrumentation


\end{keyword}

\end{frontmatter}

\section{Introduction}\label{Intro}

An apparently contradictory nature seems to oppose the field of astronomical giant telescopes, requiring always heavier infrastructures, to the field of photonics dealing with micron-sized components. However, the strong development that has characterized the field of optical telecommunications in the last forty years has begun to spread over the astronomy field in a recent past especially in the field of stellar interferometry. For instance, optical fibers are particularly interesting because they can be used as single-mode filters to improve the accuracy of the interferometric visibilities \citep{ForestoIO97,Perrin97conf,CHARA06}. In a near future, they will be implemented for light transport to interconnect telescopes at distances of a few hundred meters and to replace complex optical trains \citep{Perrin06}. Integrated optics (IO) devices are also promising for multi-aperture  interferometry. Etching or lithography techniques used in microelectronics allow us to design and to manufacture complex optical functions that combine several beams within a small and compact chip \citep{Berger01,IOTA06}. This is the prelude to a valuable miniaturization of the major components of an astronomical instrument: developing integrated optics spectrographs could be the next ambitious milestone \citep{Bland06}.\\
Photonics devices could also be very valuable in the field of space-based instrumentation where simple designs, compact and stable components with regard to external constrains (mechanical stress, temperature changes) are requested. Such an option is seriously considered to be part of space interferometry missions like {\it Darwin} and/or {\it TPF} \citep{Beichman,Fridlund}. 
However, most of the prolific exchange between astronomy and photonics occurred in the near infrared bands J (1.25 $\mu$m), H (1.65 $\mu$m) and K (2.2 $\mu$m). 
Reliable components in the mid-infrared domain -- i.e. beyond $\lambda$=5 $\mu$m -- are available for the past few years. Although several studies on guided optics have acquired some visibility in astronomy in the last five years, the technology is not fully mature yet to 
achieve an on-sky demonstration in the mid-infrared. The objective of our work is to overcome the technological limitations in order to reach such a demonstration.\\
\indent This paper presents a review of the main results of our technology research on mid-infrared single-mode waveguides for planar integrated optics, conducted within the Technology Research Program of the European Space Agency. Sect.~\ref{part2} presents the nulling instrumental issues on the wavefront quality and specifies the possible role of single-mode devices. Sect.~\ref{part3} exposes 
the recent achievements obtained on both dielectric and conductive waveguides. The conclusions are presented in Sect.~\ref{part4}.


\section{The advantages of single-mode waveguides in nulling interferometry}\label{part2}

\subsection{Short description of the instrumental issues}\label{subsect21}

In the astrophysical context of the characterization of exoplanets around solar-type stars, the first issue is the very high contrast between the two objects, making the direct observation of the planet very difficult. Depending on the wavelength, the flux from the hot star is 10$^{6}$ to 10$^{10}$ times higher than the one of the planet, with a more favorable contrast in the mid-infrared. This makes impossible the direct observation of the companion without an artificial attenuation of the starlight. The second issue is the angular separation. Even with a sufficient dynamic range of the measurement, the detection is still limited by the angular resolution delivered by the current single-dish telescopes.
Several studies have pointed out that the characterization of earth-like planets could be achieved with a mid-infrared nulling space-based interferometer that provides higher angular resolution. This was proposed as a driving concept by \citet{Bracewell} and \citet{Angel86}. \\
\indent Interferometry from space is extremely challenging. It requires a reliable concept of formation-flying for the telescope array and a new approach of satellite engineering since the degree of cancelation -- or rejection ratio -- is primarily dependent on the configuration of the nulling interferometer \citep{Angel97}. Beyond its theoretical limit, the interferometric null is also degraded by systematic and random errors on the fine path-length control, on the intensity balance and polarization matching between the incoming beams, on the achromaticity of the $\pi$-phase shift over the spectral band of interest, or on the fine control of the wavefront quality.
Concerning the last point, \citet{Leger95} have underlined that in order to reach a 10$^{5}$ null, the wavefront errors due to an insufficient optical quality, to micro-roughness and to the residual optical path difference (OPD) should be kept smaller than $\lambda$/2000 at 10 $\mu$m (i.e. $\lambda$/100 at 0.5 $\mu$m), while pointing errors should remain below 1/1600 of the Airy disk at 10 $\mu$m (i.e. 1/80 at 0.5 $\mu$m).\\
\indent To relax these severe constraints, \citet{Ollivier} proposed to use pinholes to be placed at the focus of each telescope, although this solution can only filter the small-scale defects of the wavefront (i.e. the high spatial frequencies).
Further improvements can be achieved using single-mode waveguides as described hereafter.

\subsection{Single-mode waveguides and technical requirements}\label{beam_combination}

\begin{figure*}[h]
\begin{minipage}{\textwidth}
\centering
\subfigure{\includegraphics[width=7.0cm]{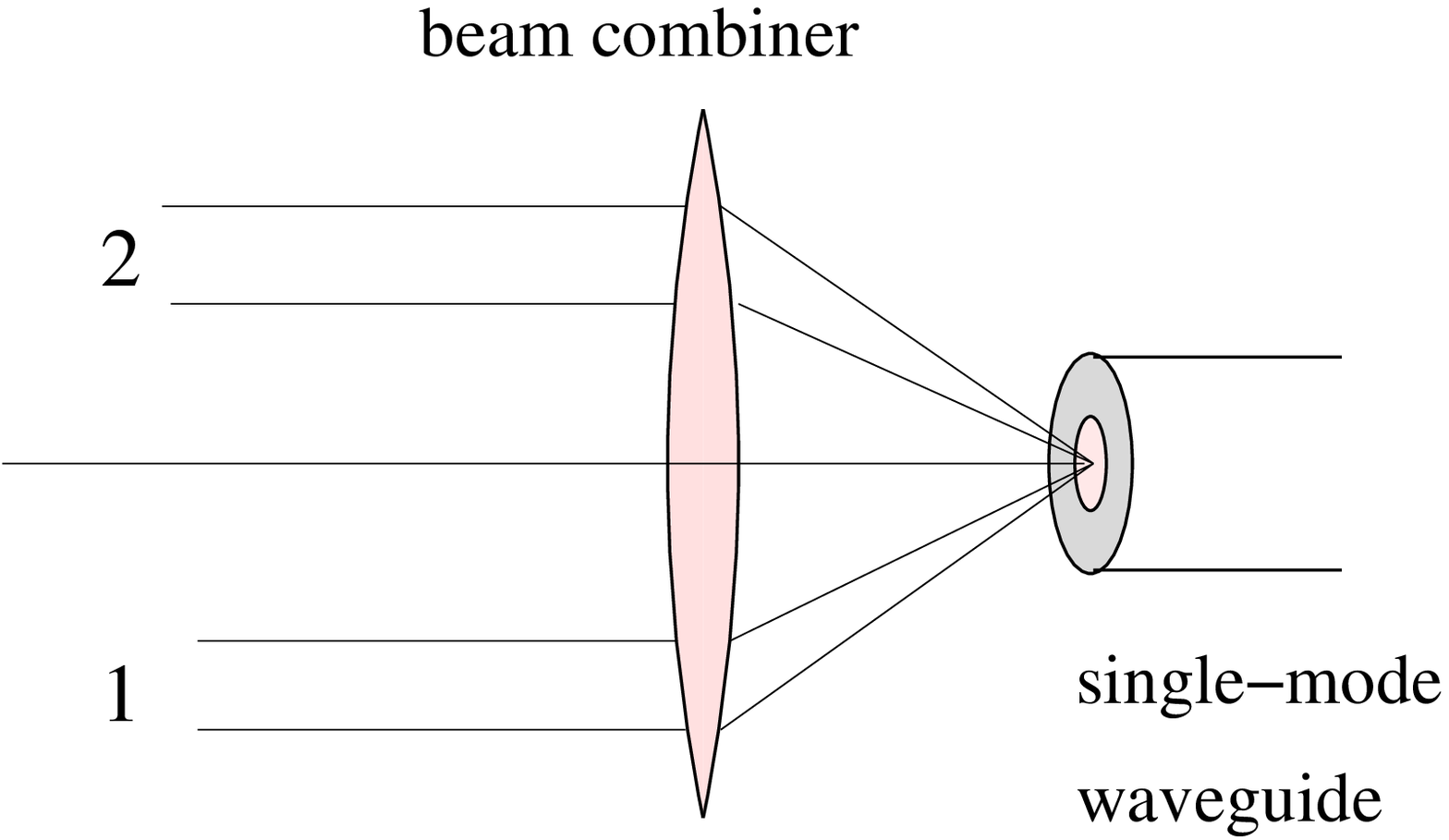}\label{fig11}}
\hspace{1.0cm}
\subfigure{\includegraphics[width=4.5cm,height=4.5cm]{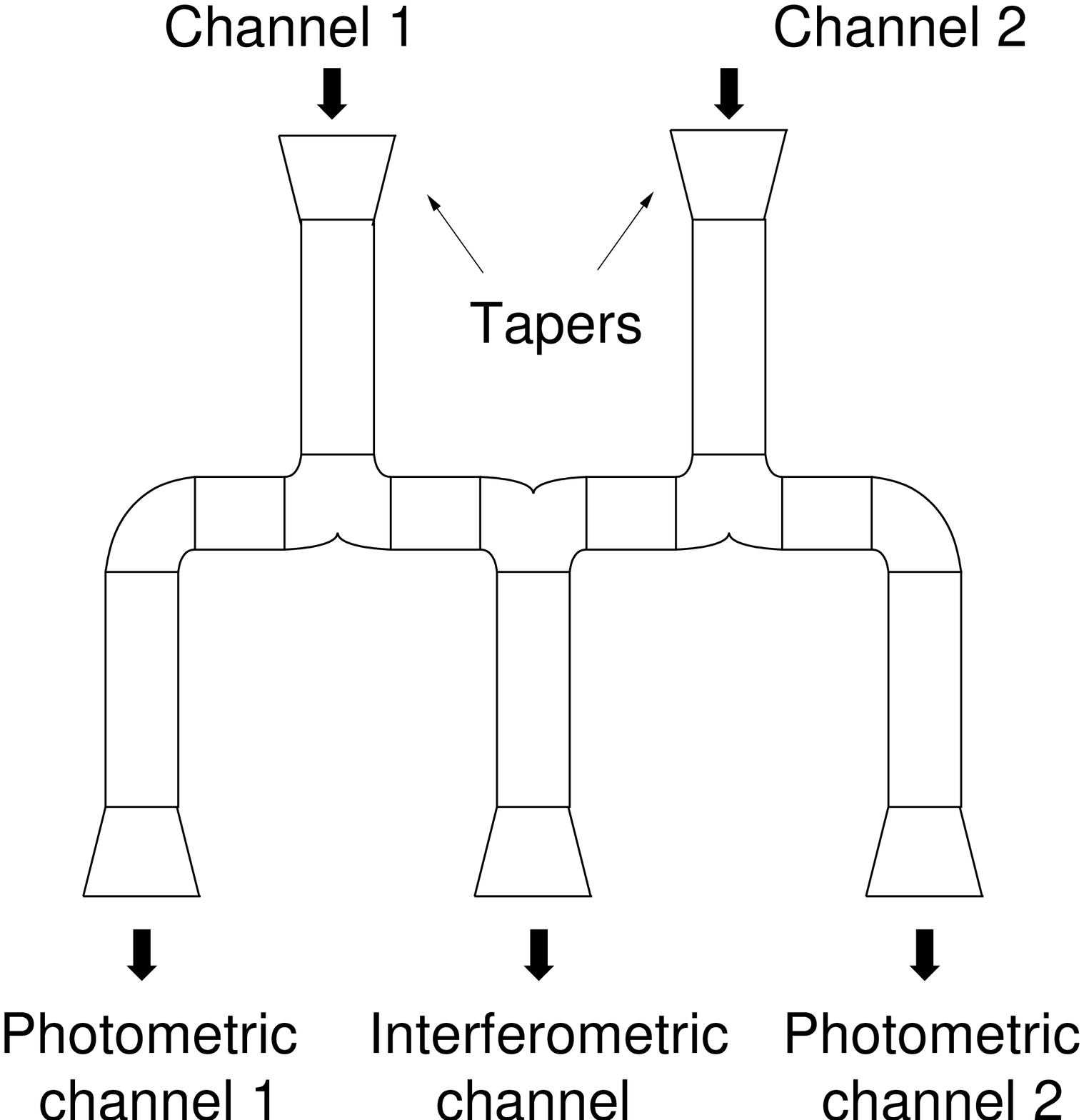}\label{fig12}}
\end{minipage}
\caption{(a): View of the multi-axial beam combination schemes using a single-mode fiber. (b): Schematic of a co-axial IO beam combiner with its two inputs and three outputs. The multi-axial design is simpler but very sensitive to beams misalignments. The co-axial design, although technologically more complex, can provide more stability with regards to external constraints.}\label{fig1}
\end{figure*}

\citet{Mennesson} have shown that the fundamental property of a single-mode waveguide -- i.e. its ability to maintain the same amplitude and phase distribution of the wavefront over the propagation distance \citep{Marcuse} -- allows to filter out the low and high order phase defects of the beam, with the exception of the residual OPD. Modal filtering can be considered either before or after the coherent beam combination, and is applicable with both the co-axial and multi-axial combination schemes (see Fig.~\ref{fig1}). 
The corollary is to benefit of reliable single-mode waveguides and/or integrated optics in the mid-infrared.
The intensity of the flux transmitted by a single-mode waveguide is directly proportional to the input coupling efficiency. This is given by the overlap integral of the product between the incident electric field and the fundamental mode distribution as:

\begin{eqnarray}
\centering
\zeta \propto \iint_{A} E_{i}.E^{\ast}_{0}dA\label{equation1}
\end{eqnarray}

\noindent where $E_{i}$ and $E_{0}$ are respectively the electric field and the fundamental mode\footnote{We do not treat here the question of the field polarization.}. The corrugation of the incoming wavefront decreases the coupling efficiency $\zeta$ -- and then the transmitted flux -- while the phase of the outing wavefront only depends on the waveguide properties. Thus, it is commonly established that a single-mode waveguide converts the phase mismatches between the wavefronts into pure intensity mismatches.

The co-axial scheme can also be implemented with guided optics devices using an integrated optics (IO) Y-junction. If the single-mode behavior of the device is preserved, the IO component combines in one single chip the functions of beam combination and modal filtering. A schematic view of an interferometric single-mode IO beam combiner is shown in Fig.~\ref{fig12}. From the technology point of view, the device is generally chemically etched on a high-index thin layer or embedded within the glass substrate by ion exchange \citep{Broquin2001}.
The basic beam combiner must have at least two inputs and three outputs. One output is the interferometric channel, the two others are the photometric channels, which are required to calibrate the intensity unbalance. The width of the waveguides before and after combination is imposed by the single-mode requirement and by the index-difference. In order to improve the coupling efficiency, tapers are generally added at the input and output of the device. A taper is a smooth transition between the geometry that matches the numerical aperture of the incoming beam, and the geometry of the channel waveguide itself. 
In the context of {\it Darwin}, a single-mode integrated optics device must fulfil several optical and thermo-mechanical requirements, which are summarized in Table~\ref{requirements}. However, since this work is the first attempt to develop an IO technology for the mid-infrared, we have considered these specifications more as a general indication than as a mandatory requirement. Furthermore, in the wider context of ground-based mid-infrared interferometry, the technical requirements would clearly be different. Consequently, we focused first on demonstrating the feasibility of thermal single-mode waveguides and on a few issues like the modal behavior and the component transmission. 

\subsection{Advances in mid-IR photonics for interferometry}

Although mid-IR photonics devices have raised interest in the medical field, the development of single-mode components has mainly been driven by needs of the astronomical field in the last years. Different teams, both in the US and in Europe, have started technology studies to manufacture efficient single-mode fibers. The first samples were produced by \citet{Borde03} using chalcogenide glass, transparent up to $\sim$15 $\mu$m.
More recently, silver halide compositions (AgClBr) have been  used to extend the transmission range towards 20 $\mu$m \citep{Wallner,Ksendzov06,Shalem05}, with reported propagation losses about 10 to 15 dB/{\it m}.\\
\indent The alternative to fiber optics is integrated optics (IO). Although the principle for field confinement remains similar, the technical implementation of integrated optics is sensitively different from fibers. In particular, the core objective of IO is to accomplish complex optical functions, which imply the mastery of a specific technology to manufacture multi-beam junctions and splitters. Thus, different time scales apply to R\&D activities for those two options. In the last years, IO solutions were investigated by a group of the University of Arizona \citep{Wehmeier} and by a consortium of laboratories in Grenoble, France, within the ESA-funded project {\it Integrated Optics for Darwin} (IODA) \citep{LabadiePhD}.

\begin{table}[t]
\centering
\begin{tabular}{|l|c |} \hline
\multicolumn{2}{|c|}{Optical requirements} \\ \hline \hline
Spectral range                     & 6 -- 20 $\mu$m   \\ \hline
Number of sub-bands                & 3 to 4 (1) \\ \hline
Transmission                       & $\ge$ 50\%  \\ \hline
Optics interface                   & f/4 numerical aperture (2) \\ \hline
Modal behavior                     & high order mode    \\
                                   & suppression to 10$^5$  \\ \hline
Interface losses (Fresnel losses)  & $\le$ 6\%  \\ \hline
Polarization control               & $\sim$ 0.1\% (3) \\ \hline
Intensity mismatch                 & $\sim$ 0.1\% (3) \\ \hline
\multicolumn{2}{|c|}{\hspace{0.5cm}\hspace{0.5cm} Thermo-mechanical requirements \hspace{0.5cm}\hspace{0.5cm}}           \\ \hline \hline
 Operating temperature & $\sim$40\,K     \\ \hline
\end{tabular}
\caption{Technical requirements for a mid-infrared IO beam combiner devoted to nulling interferometry on {\it Darwin}. (1): Sub-bands of the 6--20 $\mu$m range where the device is single-mode. (2): Numerical aperture of the beam coupled into the waveguide input. (3): The requirements on the polarization control and the intensity splitting ratio account for null leakage of 10$^{-5}$. No active control of the polarization and of the intensity mismatches is considered here.}\label{requirements}
\end{table}

The following section reports the main achievements of our project.

\section{Waveguide concepts for integrated optics: from near to mid-infrared}\label{part3}

The approach initially considered has been to extend the concept of dielectric waveguides, well mastered in the visible and near IR, to the 10-$\mu$m range using suitable transparent materials like chalcogenide glasses, zinc selenide and zinc sulfide.  A second promising approach, although less straightforward, is to use microwave engineering conductive waveguides scaled to the 10-$\mu$m range, i.e. conductive waveguides in which the confined wave is propagated by successive reflections on the metallic walls of the waveguide.

\begin{figure*}[t]
\begin{minipage}{\textwidth}
\centering
\label{mlines}
\subfigure{\includegraphics[width=4.2cm]{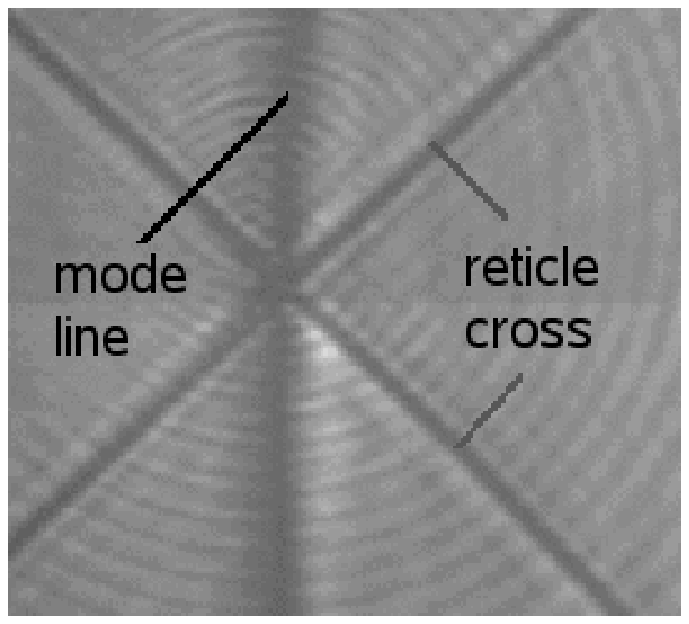}\label{mline}}
\hspace{0.2cm}
\subfigure{\includegraphics[width=6.0cm]{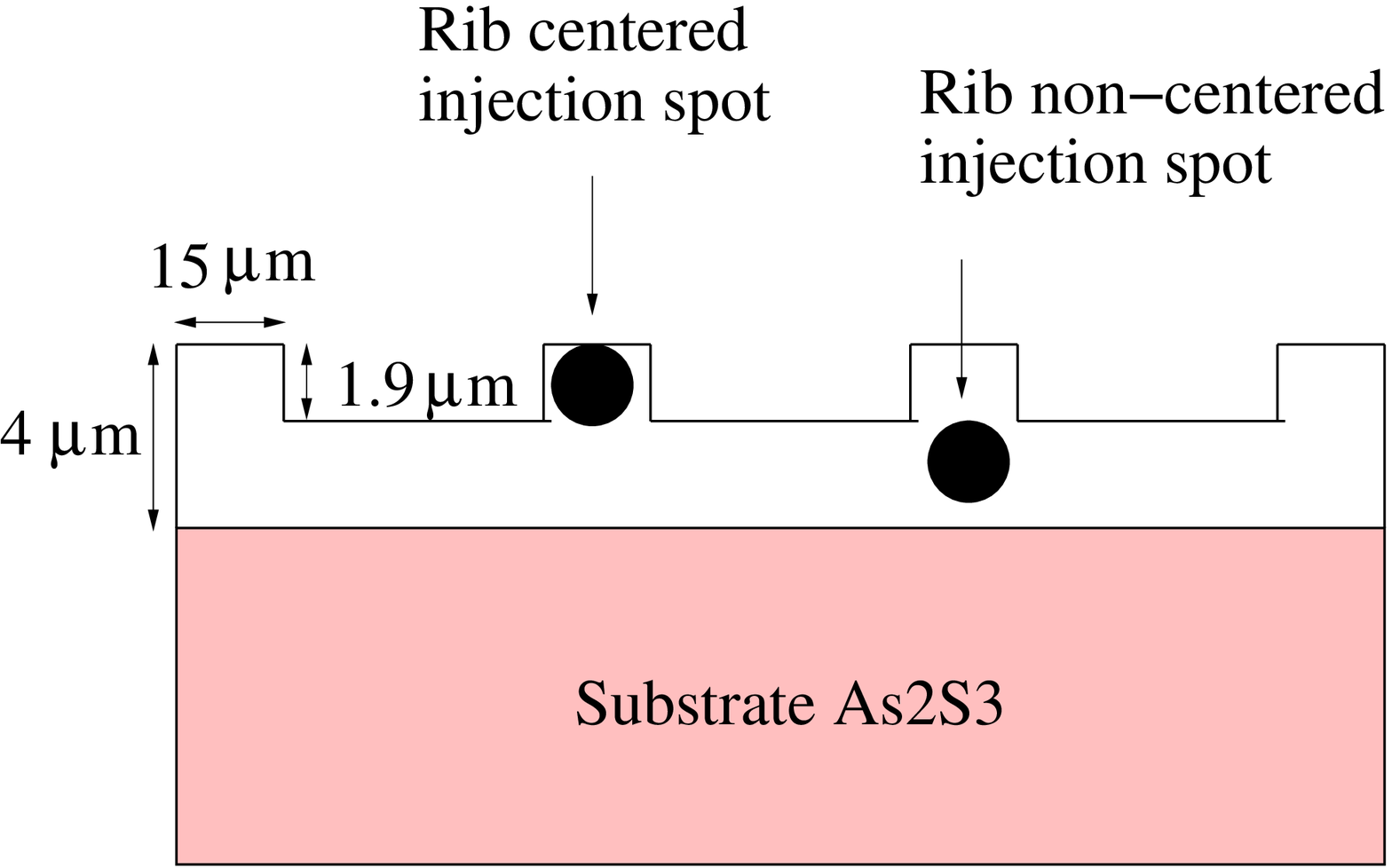}\label{ribwg1}}
\hspace{0.3cm}
\subfigure{\includegraphics[width=3.9cm]{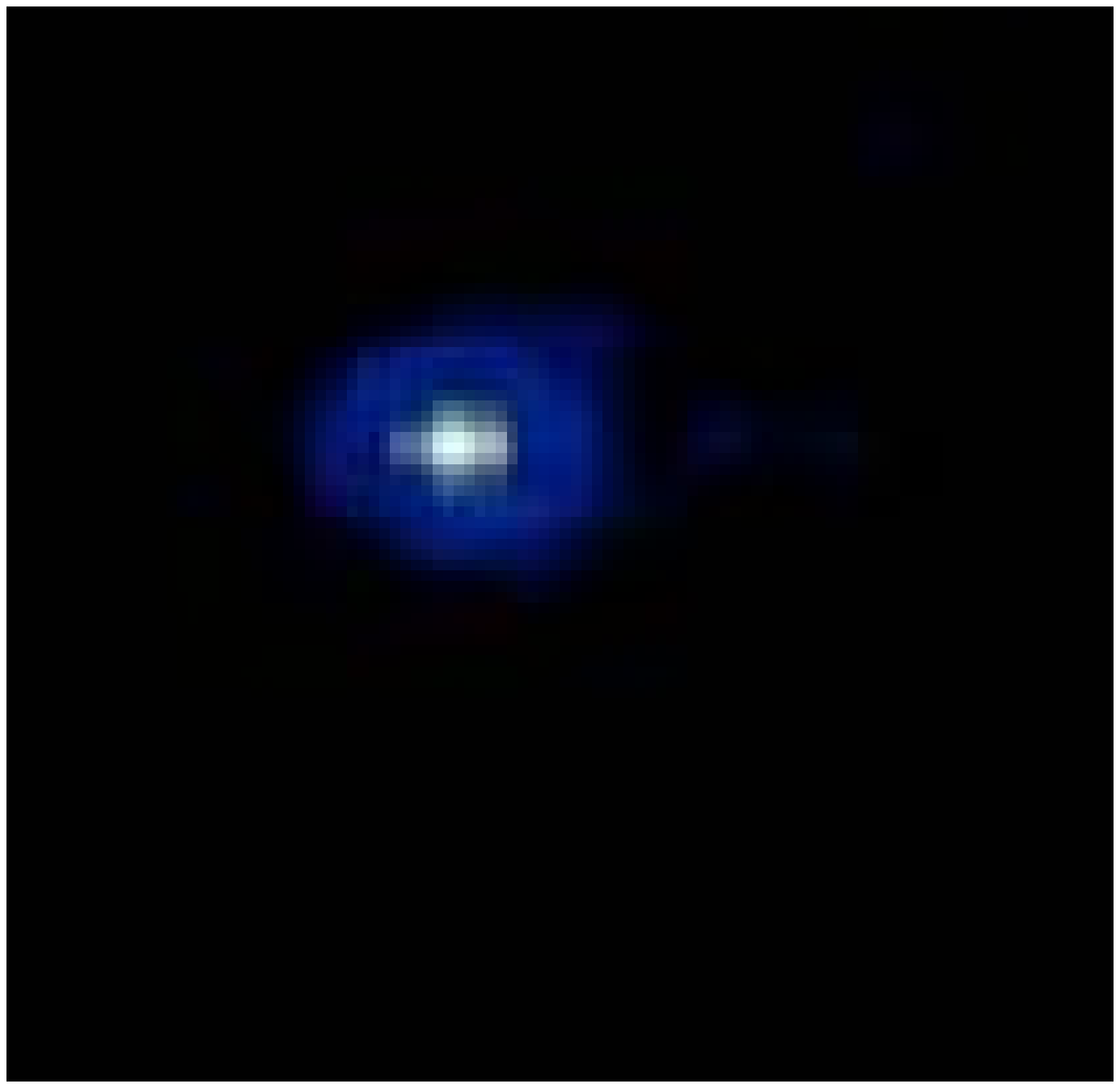}\label{ribwg2}}
\end{minipage}
\caption{(a): Output of the 10 $\mu$m {\it m}-lines experiment. The vertical dark line observed in the output beam corresponds to a {\it mode line} of the waveguide. The cross is an image of the reticle that permits to precisely point the line and to measure its angular position. (b): Schematic view of a rib-waveguide with the etched thin film deposited on a thick substrate. (c): Experimental output of a rib-waveguide for the centered-spot case. The flux is well confined within the waveguide.}\label{ribwg}
\end{figure*}

\subsection{Exploring dielectric solutions}

Dielectric waveguides (i.e. fibers, IO) are based on the confinement of an incident beam by the total reflection phenomenon 
\citep{Marcuse}. The total internal reflection occurs at the interface between a high-index core or thin film and a low-index substrate (which can also be air with $n$=1). The modal behavior of a dielectric waveguide is solely dependent on the physical dimensions of the confining media and on the refractive index difference with respect to the substrate. The waveguide is defined as single-mode when only the fundamental mode propagates and as multimode in the opposite case. The initial step has been to isolate infrared materials with suitable properties, like transmission range, chemical composition or easiness of preparation. Two compositions have been considered in this project: chalcogenide glasses on one side, zinc selenide (ZnSe) and zinc sulfide (ZnS) on the other side. Chalcogenide glass is a well-known composition, already considered for fibers production \citep{Borde03}. ZnSe and ZnS are commonly used in infrared optical components, whereas their use in the photonics field has been hardly investigated. The two types of composition present a transparency range that matches, at least partially, the 6--20 $\mu$m band with an upper limit between 15 to 20 $\mu$m \citep{Palik}. 
The planar structures have been produced by thermal evaporation of a thin film on a substrate \citep{Vigreux05, Broquin03}.\\
\indent After having identified candidate materials, the following step has been to produce slab waveguides, i.e. one-dimensional guiding structures composed by a thin high-index layer ($\sim$ few microns) evaporated on a low-index thick substrate ($\sim$2-3 mm). Two types of structures were synthesized and studied. The first one is Zinc Selenide films on a Zinc Sulfide substrate (ZnSe/ZnS). The second one is based on the deposition of selenide films (As$_{2}$Se$_{3}$) or telluride films (Te$_{2}$As$_{3}$Se$_{5}$) on a sulfide-type substrate (As$_{2}$S$_{3}$). The compositions of the film and the substrate have to be relatively close in order to limit the mechanical stress of the layer during the cooling process.
An important characterization aspect of this phase has been to assess experimentally the modal behavior of the slab waveguides, i.e. to ensure that the geometrical parameters of these new structures (film thickness, index difference) were adequate to propagate at least one mode. For this purpose, the {\it m}-lines method was used \citep{Tien70}. This method allows to couple light into the slab waveguide using a prism-coupler and to observe, in the decoupled beam, the discrete signatures of the waveguide propagating modes \citep{Tien69}. The measured quantity is the {\it spectrum of  mode indices} from which the quantities $n_{c}$ and $d$, respectively the refractive index and the thickness of the layer, are derived by solving the dispersion equation of the waveguide.
\noindent We implemented the {\it m}-lines method in the mid-infrared \citep{LabadiePhD} to investigate the modal behavior of several slab structures. For both selenide and telluride compositions, two to three mode lines were experimentally observed at 10.6 $\mu$m, suggesting the multimode feature of the slab. 
An example of the output of the 10 $\mu$m {\it m}-lines experiment is shown in Fig.~\ref{mline}. The multimode behavior allows to remove the degeneracy of the solution ($n_{c}$, $d$) encountered when only one mode index is measured. This method permitted us to measure the refractive index of Arsenic-based and Telluride-based films with an accuracy of $\pm$0.001 \citep{LabadieOpEx}. We found that the two compositions have a mean index $n_{c}$$\backsimeq$2.7, which is appreciably higher than the As$_{\rm 2}$S$_{\rm 3}$ substrate index at the same wavelength ($n_{c}$$\backsimeq$2.4).\\
\indent A similar characterization was conducted at 10.6 $\mu$m on the ZnSe/ZnS composition, which resulted in the observation of only one mode line in the accessible angle range. This result suggested that the  opto-physical parameters of the structure allow only one mode to propagate in the cavity. This result has been confirmed first by solving the degeneracy for the quantity $d$ at a shorter wavelength ($\lambda$=1.2 $\mu$m, i.e. where the structure is highly multimode) and secondly deriving $n_{c}$ at 10.6 $\mu$m.  The {\it m}-lines data reduction resulted into a refractive index $n_{c}$$\backsimeq$2.45 and a film thickness $d$$\backsimeq$6 $\mu$m. 

\begin{figure*}[t]
\begin{minipage}{\textwidth}
\centering
\subfigure{\includegraphics[width=5.0cm]{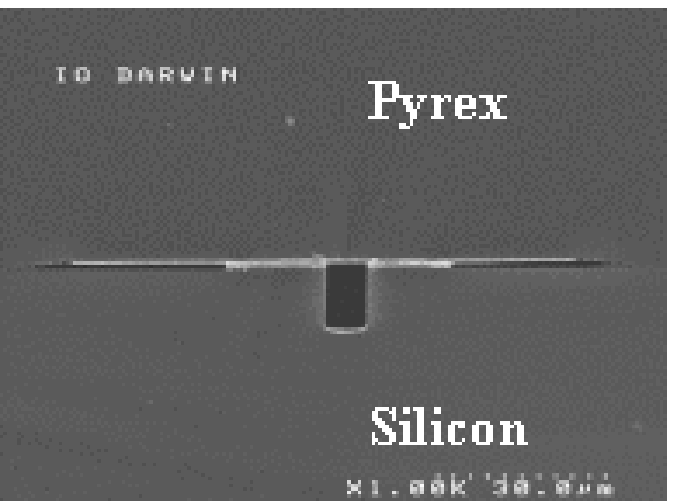}\label{chip1}}
\hspace{1.0cm}
\subfigure{\includegraphics[width=5.7cm]{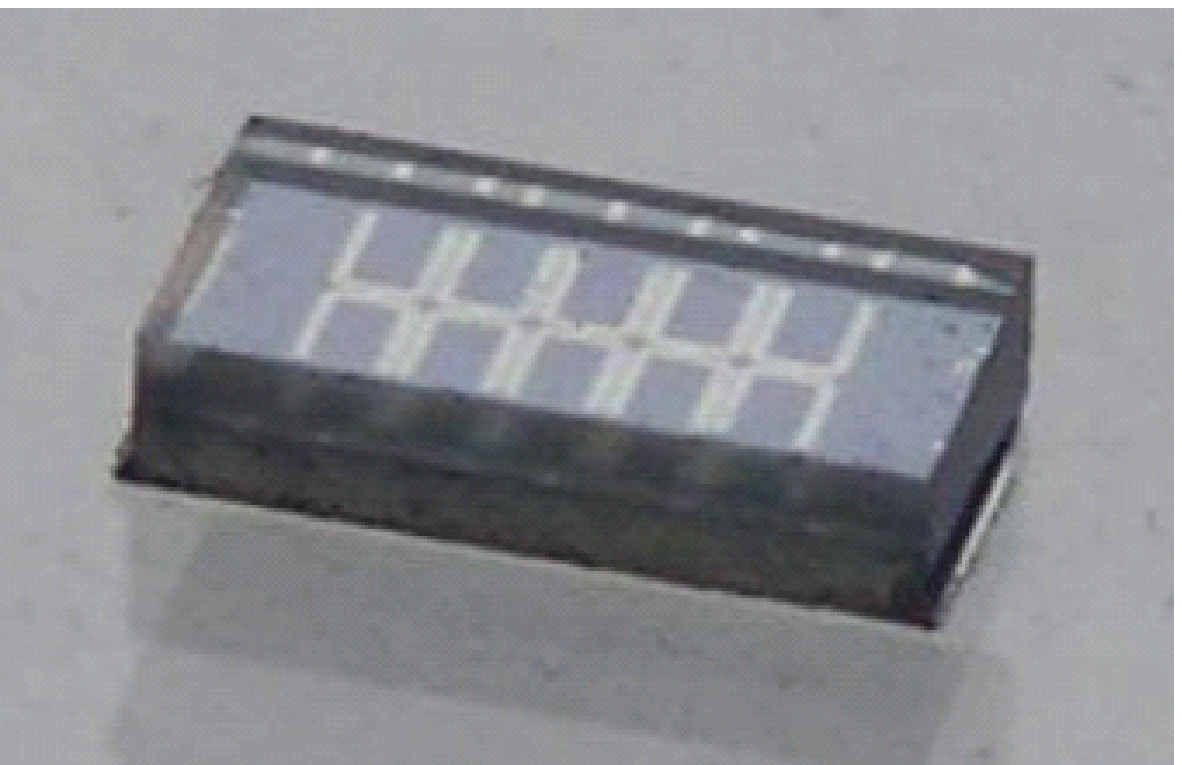}\label{chip2}}
\end{minipage}
\caption{(a): View of the output of a HMW at the scanning electron microscope. The smaller side of the waveguide is 5 $\mu$m, the larger side is 10 $\mu$m. (b): Picture of a chip containing hollow metallic waveguides. The sample is 5 mm large, 2 mm long, 1 mm width. \newline}\label{hmw1}
\end{figure*}

The experimental result permitted us to verify the effective single-mode behavior of the ZnSe/ZnS structures using a simple model of slab waveguide \citep{LabadiePrep}.\\
\indent The previous characterization phase, which demonstrated the feasibility of mid-infrared planar waveguides, has been the mandatory prerequisite to undergo the manufacturing process of channel waveguides.\\
\indent An additional step towards the planar integrated optics has been to manufacture a first channel waveguide in order to observe the confinement of the field. The first attempt has been achieved so far only with chalcogenide planar waveguides for which a sufficient reliable chemical etching process of the layer has been developed. The option of ZnSe/ZnS is not considered yet in the following discussion. The process undertaken has been the manufacturing of rib-waveguides based on a reactive ion etching of a Te$_{2}$As$_{3}$Se$_{5}$ film deposited on an As$_{2}$S$_{3}$ substrate under CF$_{4}$/O$_{2}$ atmosphere. The general shape and the dimensions of the rib-waveguide is illustrated in Fig.~\ref{ribwg1} (not to scale). The sample has undergone characterizion tests to assess the confinement of the field along the propagation length. For these tests, we used the injection bench \citep{LabadiePhD} at 10.6 $\mu$m to couple infrared light at the waveguide input. The 10.6 $\mu$m image clearly shows a good lateral confinement of the field when the injection spot is centered on the rib itself, as presented in Fig.~\ref{ribwg2}. 
As the injection spot is translated vertically below the rib, the lateral confinement of the field is lost and a horizontal strip is observed at the waveguide output. 
The propagation losses for this channel waveguide were roughly estimated at 10 dB/cm \citep{Vigreux07}.

\subsection{Introducing conductive waveguides}

In this approach, we investigated the possibility to scale to the 10 $\mu$m range the concept of conductive -- or metallic -- waveguides, which are widely used in the millimeter range for radioastronomy. Conductive waveguides, also named hollow metallic waveguides (or HMW), have been extensively studied in microwave engineering. The evolution towards shorter infrared wavelengths has been limited for some time by the ability to produce small devices at the 10-20 $\mu$m scale. The continuous improvement of micro-machining tools permits now to overcome this barrier \citep{AubignyPhD}. The advantage of conductive waveguides for nulling interferometry has already been underlined theoretically by \citet{Wehmeier}. Once again, our goal has focused on demonstrating experimentally that single-mode conductive waveguides could effectively be manufactured and operated at 10 $\mu$m.\\
\indent In a previous paper \citep{LabadieAA}, we demonstrate that a suitable geometry for conductive waveguides is a rectangular shape with dimensions $a$ and $b$, where $a$=2$b$. Under this condition, the spectral range where only the fundamental mode TE$_{10}$ propagates is $a$ $<$ $\lambda$ $<$2$a$. \\
\indent The manufacturing process used for the fabrication of HMW is a standard process of silicon substrate etching followed by an anodic bonding of a Pyrex cover. The inside walls are coated with a gold layer having a thickness greater than the 50 nm skin depth of the metal at 10 $\mu$m. The wave is propagated by reflections on the walls in the internal dielectric medium which is air. Fig.~\ref{chip1} shows a view of a rectangular HMW with dimensions $a$=10 $\mu$m and $b$=5 $\mu$m. The etching accuracy is better than 100 nm. The wide horizontal pattern shows the boundary between the silicon substrate, in which the waveguide has been etched, and the Pyrex cover. The four lateral walls have been coated with gold before the bonding process between the substrate and the cover. Fig.~\ref{chip2} shows a typical chip containing the conductive waveguide. The chip is 5-mm long, 1-mm width and 1-mm height. The waveguides are aligned perpendicularly to the longer side of the chip and thus they have also a length of 1-mm too.\\
\indent When possible, one-dimensional tapers have been added to the inputs and outputs of the waveguides to increase the coupling efficiency and to reduce the impedance mismatch. This effect has been quantified to $\sim$50\% of coupling improvement with respect to the case where no tapers are added \citep{LabadieAA}.

\begin{figure*}[t]
\begin{minipage}{\textwidth}
\centering
\subfigure{\includegraphics[width=5.2cm]{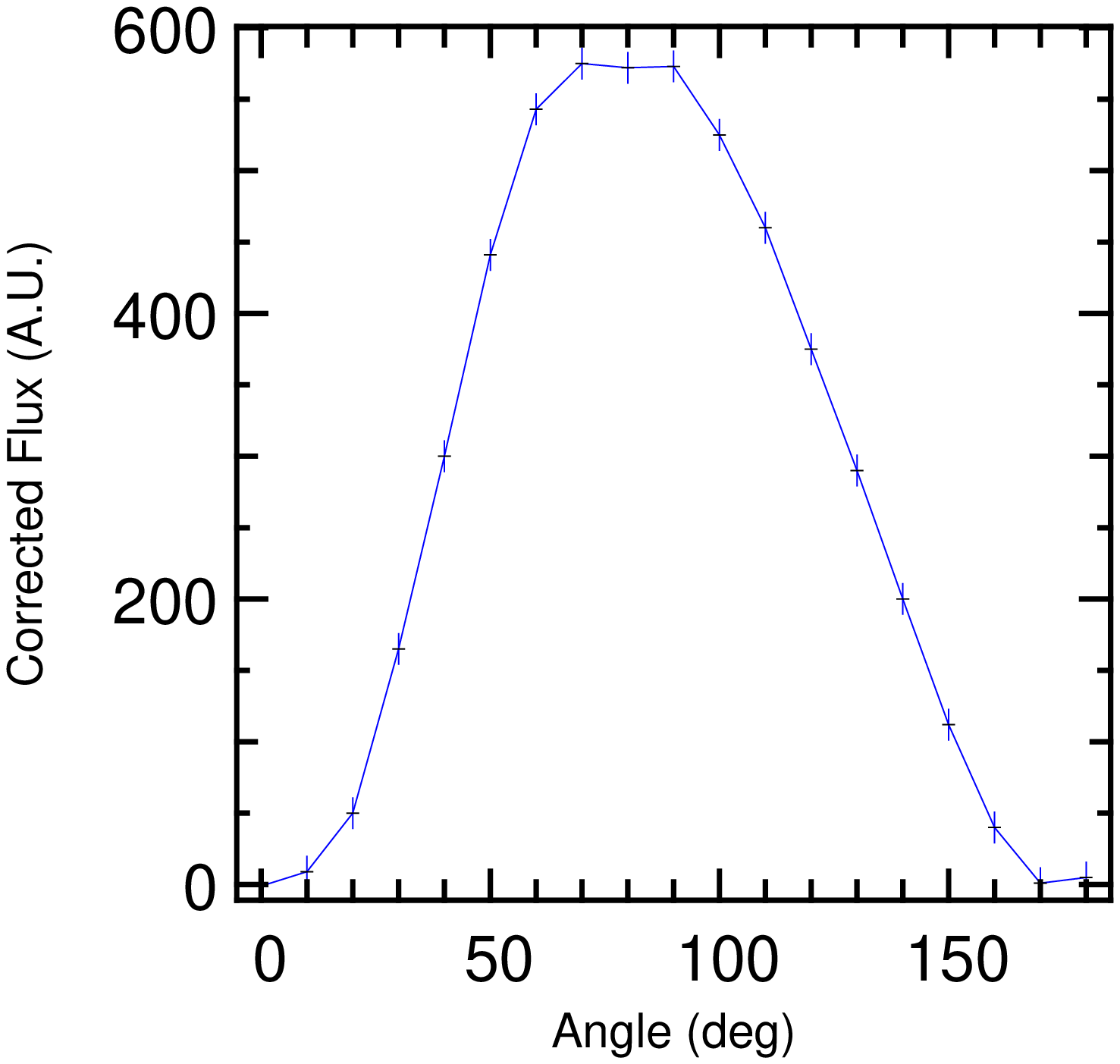}\label{curve1}}
\hspace{0.8cm}
\subfigure{\includegraphics[width=5.3cm]{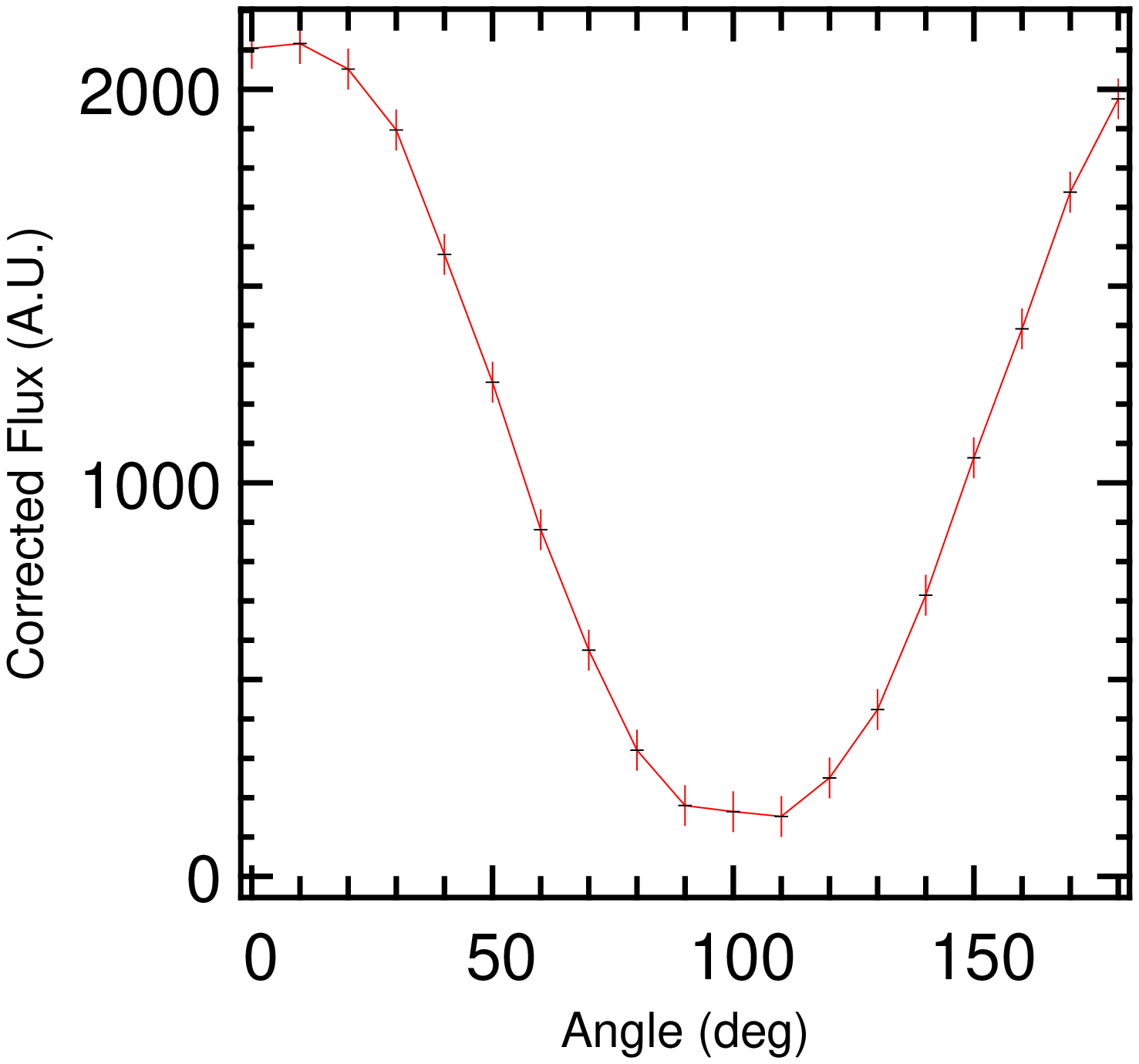}\label{curve2}}
\end{minipage}
\caption{Transmitted flux as a function of the polarization angle of the exiting wave. (a): The single-mode case with $b_{\rm 1}$=5 $\mu$m shows a clear extinction at symmetric positions 0$^{\circ}$ and 180$^{\circ}$. (b): for a multimode waveguide with $b_{\rm 2}$=9 $\mu$m, no total extinction is achievable. The graphs are from \citet{LabadieAA}.}\label{polarization}
\end{figure*}

The characterization of the modal behavior of the samples at infrared wavelengths has been the fundamental aspect of our research on conductive waveguides. Differently from the dielectric option, the manufacture process itself is sufficiently mature to provide theoretical first single-mode sample, at least "on the paper". The experimental verification of such a statement is based on the analysis of the polarization of the wave exiting the waveguide. Indeed, the fundamental mode TE$_{10}$ propagated by a single-mode rectangular HMW presents a linear polarization perpendicular to the long side of the waveguide. Oppositely, a multimode structure will propagate at least two modes, TE$_{10}$ and TE$_{01}$, which present a perpendicular linear polarization with respect to each other. By placing a grid polarizer after the waveguide in the injection setup, it is possible to measure the transmitted flux as a function of the analyzed polarization angle. This experiment was carried out on two different samples with $a$=10 $\mu$m and $b_{1}$=5 $\mu$m and $b_{2}$=9 $\mu$m respectively. At an operating wavelength of 10.6 $\mu$m, a single-mode behavior is expected for the first sample and a multimode behavior for the second sample. Fig.~\ref{polarization} compares the experimental transmission curves obtained in the two cases described above. The light entering the waveguide has an elliptical polarization thanks to a Cadmium Sulfide (CdS) quarter-wave plate added in the optical path so that all the polarization directions are probed. The plot of Fig.~\ref{curve1} shows the experimental curve of an HMW expected to be single-mode by design at 10 $\mu$m. We observe that for the two angular directions 0$^{\circ}$ and 180$^{\circ}$ the flux is extinguished to the noise level, while the transmission is maximum for 90$^{\circ}$, which corresponds to the angular direction of the electric field. The plot of Fig.~\ref{curve2} corresponds to the experimental case of a HMW expected to be multimode at 10 $\mu$m. It indicates that no extinction is observed for any angular direction. This is due to the combined contribution of two perpendicular polarizations in the cavity, as expected for a multimode structure. This result represents the first experimental observation of the single-mode behavior of a conductive waveguide at 10 $\mu$m.

\begin{figure}[b]
\centering
\includegraphics[width=7.0cm]{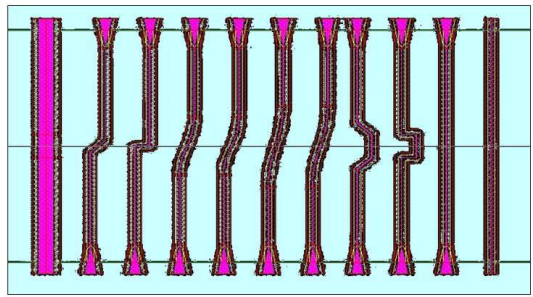}\label{curve}
\caption{Layout of a chip containing curved HMW. Different radius of curvature were tested. The sample is 5 mm large, and 1 mm or 2 mm long depending on the chip.}\label{ioda2}
\end{figure}

\indent The preliminary measurements of our single-mode HMW excess losses revealed a value of $\sim$ 4 dB/mm. In a recent work \citep{Tiberini07}, the authors have deepened the transmission analysis including the study of curved waveguides (see Fig.~\ref{ioda2}). These waveguides are the intermediate step to be assessed in order to evolve from channel waveguides to an IO beam combiner. The excess losses of the channel waveguides have been reevaluated to $\sim$ 15 dB/mm, underlying that they are strongly dependent on the size of the single-mode waveguide and on the roughness of the metallic coating. Concerning the curved waveguides, the measurement of the total throughput has permitted us to rigorously estimate the effect on the transmission of different radii \,of curvature. For waveguides smoothly bent (i.e. waveguides \#\,4,5,6,7 starting from left on Fig.~\ref{ioda2}), the measured total throughput is of the same order as for the channel waveguides. For waveguides with a more abrupt curvature (e.g. \#\,2,3,8,9), the transmission drops significantly and the output signal becomes barely detectable with the current setup.

\indent This result implies that a potential future IO combiner based on conductive waveguides will have to support large radii \,of curvature, which increases the total length of the device to more than the current 1-mm length. The resulting drawback is a higher attenuation of the transmitted flux. As a consequence, it is still required to understand whether the difference from the theoretical models of \citet{Wehmeier} is intrinsic to the technology of conductive waveguides or if unexpected defects of the current technological run (e.g. an increase of the roughness of the metallic coatings in the waveguide) are responsible of a degraded transmission.\\
\indent As part of the characterization phase, we also tested the requirement on the operating temperature for the conductive waveguides (see Table~\ref{requirements}). We dropped a single-mode component into the liquid nitrogen at 77\,$^{\circ}$K for few minutes in order to test its resistance. Afterwards, we were still able to reproduce the polarization measurements with no observable modification. This helps us to make a conclude that the components are resistant to low temperature.

\section{Conclusions}\label{part4}

The process that brings to mid-infrared integrated optics involves a technology development plan which is fairly different from the fibers one. We have presented in this paper two possible directions to reach this objective.\\
\indent Developing planar integrated optics has certainly the advantage of a good transparency in the mid-infrared and has reached a certain degree of maturity from the point-of-view of the material processing. However, the channel waveguides characterized so far are not yet single-mode and a refinement of the etching process might be necessary to better constrain the waveguides dimensions.\\ 
\indent On the other side, first single-mode conductive waveguides have been successfully characterized in the mid-infrared, although they present relatively high propagation losses. This aspect, which is a drawback for IO functions longer than 1-mm, can be minimized by shortening the waveguide and implementing large tapers in order to improve the coupling efficiency and preserve sufficient handling. Under those conditions, a single-mode HMW can be used as an efficient modal filter. Indeed, recent mid-infrared interferometric measurements have shown that nulling ratios of 10$^{4}$ or better can be obtained using a single-mode conductive waveguide \citep{LabadieAA2}.\\
\indent In the future, we suggest to carefully maintain the development of dielectric planar IO, which relies mainly on stabilization over time. In addition, a strengthening aspect would be the addition of a highly filtering conductive waveguide to the end part of a dielectric IO Y-junction.








\bibliography{report}   
\bibliographystyle{elsart-harv}

\end{document}